%% file: gersabeck_charm_cpv_lhcb.tex
\documentclass[12pt]{article}
\usepackage{graphicx}

\usepackage{hepnames}

\newcommand\pubnumber{}
\newcommand\pubdate{\today}

\def\cern{CERN, CH-1211 Geneva, Switzerland}

\def\Title#1{\begin{center} {\Large #1 } \end{center}}
\def\Author#1{\begin{center}{ \sc #1} \end{center}}
\def\Address#1{\begin{center}{ \it #1} \end{center}}

\newcommand\pubblock{\rightline{\begin{tabular}{l} \pubnumber\\
         \pubdate  \end{tabular}}}
\newenvironment{Abstract}{\begin{quotation}  }{\end{quotation}}
\newenvironment{Presented}{\begin{quotation} \begin{center} 
             PRESENTED AT\end{center}\bigskip 
      \begin{center}\begin{large}}{\end{large}\end{center} \end{quotation}}

\input econfmacros.tex

\newcommand{\agamma}{\ensuremath{A_{\Gamma}}}
\newcommand{\ycp}{\ensuremath{y_{CP}}}

\newcommand{\pbinv}{\ensuremath{{\rm pb}^{-1}}}
\newcommand{\nbinv}{\ensuremath{{\rm nb}^{-1}}}
\newcommand{\tev}{\ensuremath{{\rm TeV}}}
\newcommand{\gevc}{\ensuremath{{\rm GeV}/c}}
\newcommand{\mhz}{\ensuremath{{\rm MHz}}}

\begin{document}
\begin{titlepage}
\pubblock

\vfill
\Title{Searches for CP Violation in Charm Mixing at LHCb}
\vfill
\Author{Marco Gersabeck\\ on behalf of the LHCb Collaboration}
\Address{\cern}
\vfill
\begin{Abstract}
LHCb has started its charm physics programme using the data taken at the LHC.
The first measurements of open charm production cross-sections for proton proton collisions at $\sqrt{s}=7~\tev$ are presented.
The cross-sections of the \PDzero, $\PDstar^+$, \PDplus, and \PDsplus mesons are found to be in broad agreement with theory predictions.
The prospects for measurements of charm mixing, CP and T violation in decays of neutral \PD mesons at LHCb are discussed.
Furthermore, plans for CP violation measurements using charged \PD mesons are presented.
Most analyses are expected to yield results improving the current world averages based on the data expected to be taken in 2011.
\end{Abstract}
\vfill
\begin{Presented}
CKM2010, the 6th International Workshop on the CKM Unitarity Triangle\\
University of Warwick, UK, 6-10 September 2010
\end{Presented}
\vfill
\end{titlepage}
\def\thefootnote{\fnsymbol{footnote}}
\setcounter{footnote}{0}

\section{Introduction}
Recent charm physics precision measurements have yielded excellent results from first evidence for charm mixing~\cite{bib:charm_mixing} to searches for CP violation~\cite{bib:charm_cpv}.
LHCb will add a new chapter to this field making use of the high charm cross-section and the detector which is optimised for precision heavy flavour physics.
The LHCb experiment will be introduced in the following section.
Section~\ref{sec:xsec} presents the first measurement of charm production cross-sections at centre of mass energy of $7~\tev$.
Section~\ref{sec:mixing} covers the prospects for charm mixing measurements at LHCb followed by a discussion of charm CP violation measurements (Sec.~\ref{sec:cpvmixing}) and conclusions in section~\ref{sec:conclusion}.

\subsection{The LHCb Experiment}
The LHCb experiment~\cite{bib:LHCb} at the LHC is specifically aimed at making precision measurements with heavy flavour particle decays.
It has been built as a single forward arm spectrometer covering an acceptance of roughly $15~{\rm mrad}$ to $300~{\rm mrad}$, corresponding to $1.9$ to $4.9$ in units of pseudo-rapidity.
This design follows the angular distribution of heavy flavour quark-anti-quark pairs.
The pairs are predominantly produced in the same direction and close to the direction of one of the beams.

The LHCb detector comprises a tracking system which consists of a silicon strip VErtex LOcator (VELO), a $4~{\rm Tm}$ dipole magnet with one silicon strip tracking station before and three stations after the magnet.
The tracking stations after the magnet have silicon strip detectors in the high occupancy inner region close to the beam pipe and straw tracker modules away from the beam pipe to complete the acceptance coverage.

The LHCb detector is completed by a number of particle identification devices.
Two Ring Imaging CHerenkov detectors with a total of three different radiators allow excellent separation of pions, kaons and protons over a momentum range from $2~{\rm GeV}/c$ to above $100~{\rm GeV}/c$.
This allows measurements of hadronic decays with very high purities.
The calorimetry detectors comprise a scintillating pad detector for fast information in the trigger and for electron identification, a preshower detector, an electromagnetic and a hadronic calorimeter.
The third set of particle identification devices are five muon stations, one of which is located upstream of the calorimeters to aid the tracking of muons and four stations downstream of the calorimeter.

The LHCb trigger system uses a hardware-based system with partial readout of the detector to reduce the event rate to a maximum of $1~\mhz$ followed by a two stage software trigger based on the full detector readout.
The trigger is optimised for selecting \PB hadron decays and hence favours the charm decays originating from \PB decays which have higher transverse momentum.
The data acquired by LHCb at the time of the conference were over $3~\pbinv$ in integrated luminosity taken with an efficiency above $90\%$.

\section{Charm Production Cross-Sections}
\label{sec:xsec}
At LHCb the production of charm particles can be measured at an unprecedented centre of mass energy as well as down to a transverse momentum of $0~\gevc$.
The measurements presented here are based on a data sample of $1.8\nbinv$ of integrated luminosity.
This sample was taken with trivial trigger conditions which are assumed to be fully efficient.
This assumption has been confirmed by simulation as well as comparisons of the triggers among each other.

In principle LHCb has access to all open charm hadrons.
However, this early measurement only includes the following decays: $\PDzero\to\PKminus\Ppiplus$, $\PDstar^+\to\PDzero(\PKminus\Ppiplus)\Ppiplus$, $\PDplus\to\PKminus\Ppiplus\Ppiplus$, $\PDsplus\to\Pphi(\PKplus\PKminus)\Ppiplus$ (and charge conjugates).

The measurements were performed by extracting the raw event yields from fits to the invariant mass distributions in bins of rapidity and transverse momentum.
The contamination from charm originating from decays of heavier, long lived particles was estimated by fits to the distribution of the \PD impact parameter with respect to the event primary vertex.
This is zero (apart from resolution effects) for \PD originating from the interaction vertex, but in general takes larger values if the \PD is the daughter of a long lived particle.

The efficiency of the reconstruction and selection has been estimated using Monte Carlo (MC) simulation events.
Systematic uncertainties were obtained from data-MC comparisons or from purely data-driven methods.
The uncertainty on the tracking efficiency as estimated on data is a dominant contribution with fully correlated uncertainties of $3\%$ for each high momentum track and $4\%$ for the softer pion of the $\PDstar^+\to\PDzero\Ppiplus$ decay.
The efficiency of cuts on particle identification variables has been fully estimated on data using pure control samples.

The integrated luminosity was measured by LHCb directly using measurements of the beam sizes and overlaps with the VELO and beam current measurements from the LHC~\cite{bib:LHCb_KShort}.
The relative uncertainty of the luminosity is $10\%$ which is dominated by the beam current measurement.
The last ingredient to the cross-section measurement is the branching fraction for which the current world average has been used~\cite{bib:PDG}.

\begin{figure}[htb]
\begin{minipage}{0.49\textwidth}
\centering
\includegraphics[width=\textwidth]{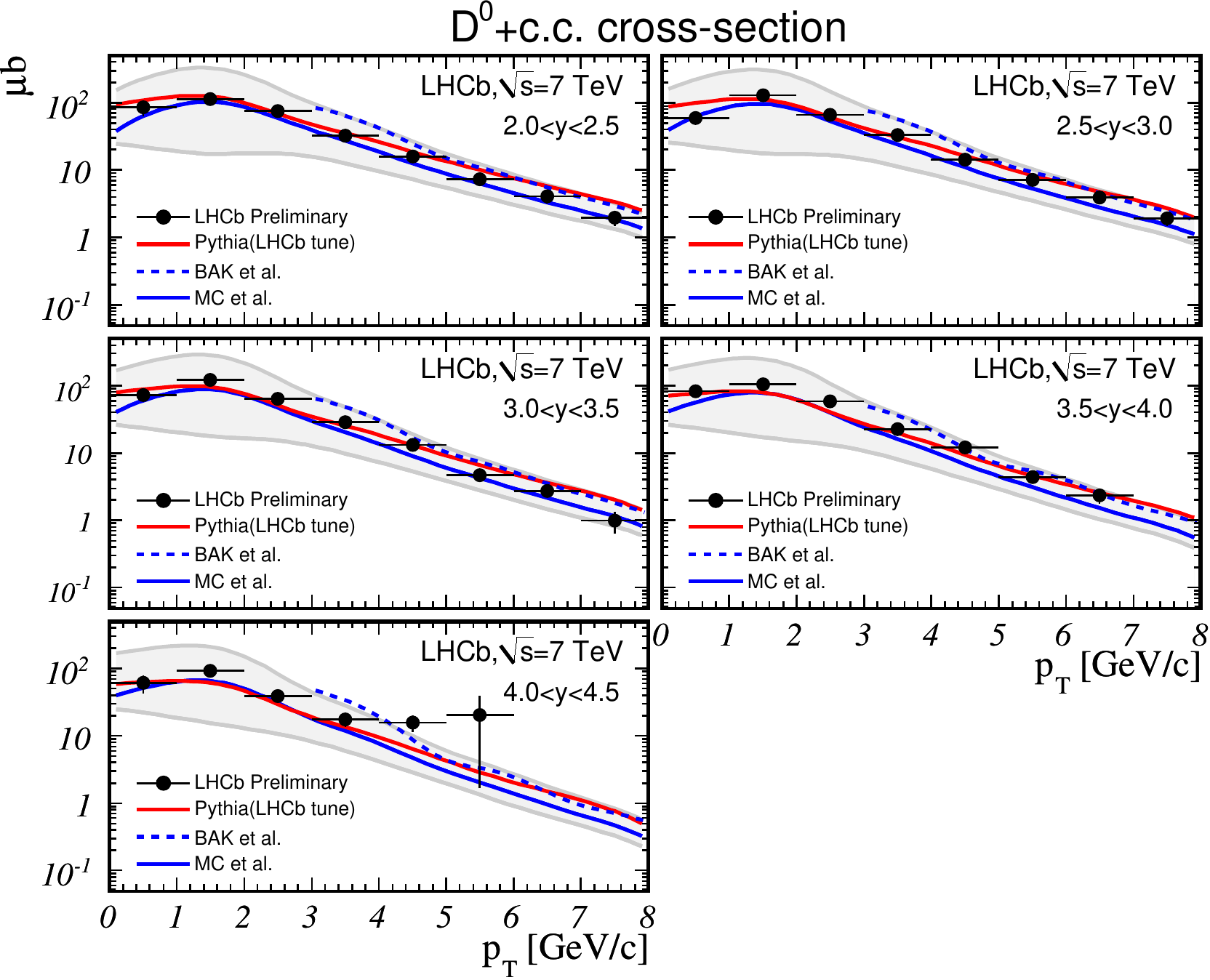}
\caption{Measured \PDzero cross-sections compared to theoretical prediction. Data are shown as a function of transverse momentum for different ranges in rapidity. The error bars show statistical and uncorrelated systematics added in quadrature. In addition there is a global correlated error of $12\%$.}
\label{fig:xsec_dzero}
\end{minipage}
\begin{minipage}{0.49\textwidth}
\centering
\includegraphics[width=\textwidth]{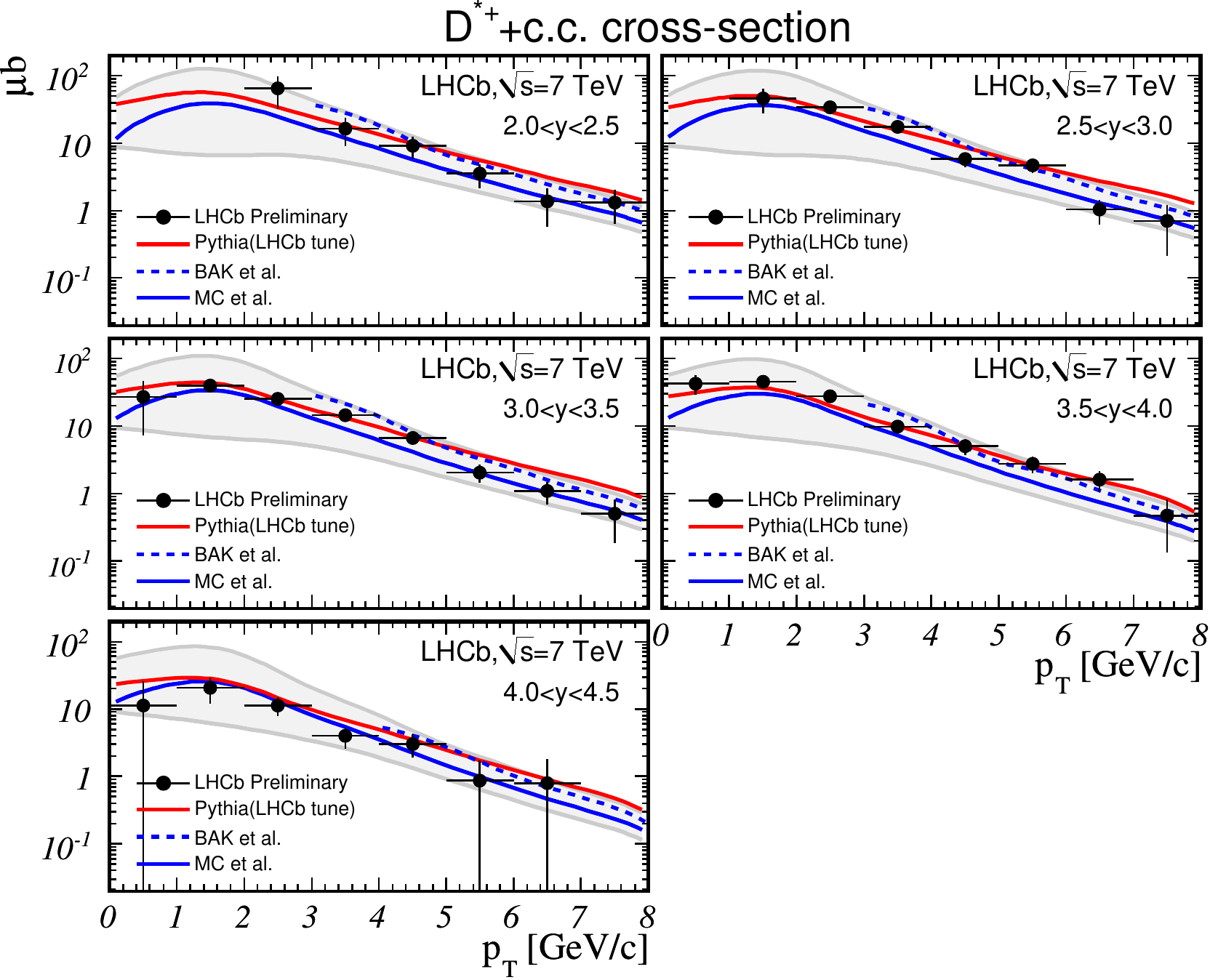}
\caption{Measured $\PDstar^+$ cross-sections compared to theoretical prediction. Data are shown as a function of transverse momentum for different ranges in rapidity. The error bars show statistical and uncorrelated systematics added in quadrature. In addition there is a global correlated error of $14\%$.}
\label{fig:xsec_dstar}
\end{minipage}
\begin{minipage}{0.49\textwidth}
\centering
\includegraphics[width=\textwidth]{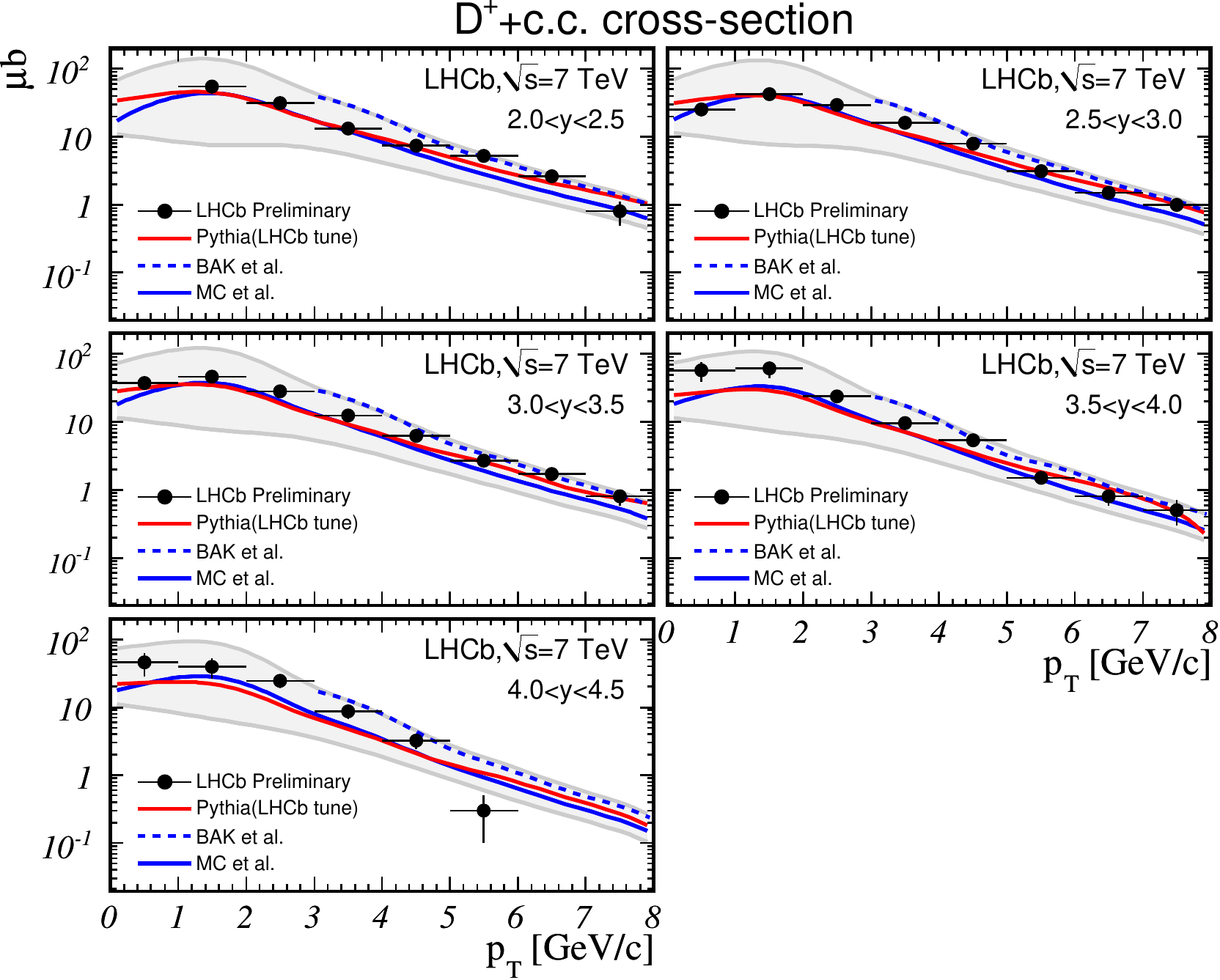}
\caption{Measured \PDplus cross-sections compared to theoretical prediction. Data are shown as a function of transverse momentum for different ranges in rapidity. The error bars show statistical and uncorrelated systematics added in quadrature. In addition there is a global correlated error of $14\%$.}
\label{fig:xsec_dplus}
\end{minipage}
\begin{minipage}{0.49\textwidth}
\centering
\includegraphics[width=0.67\textwidth]{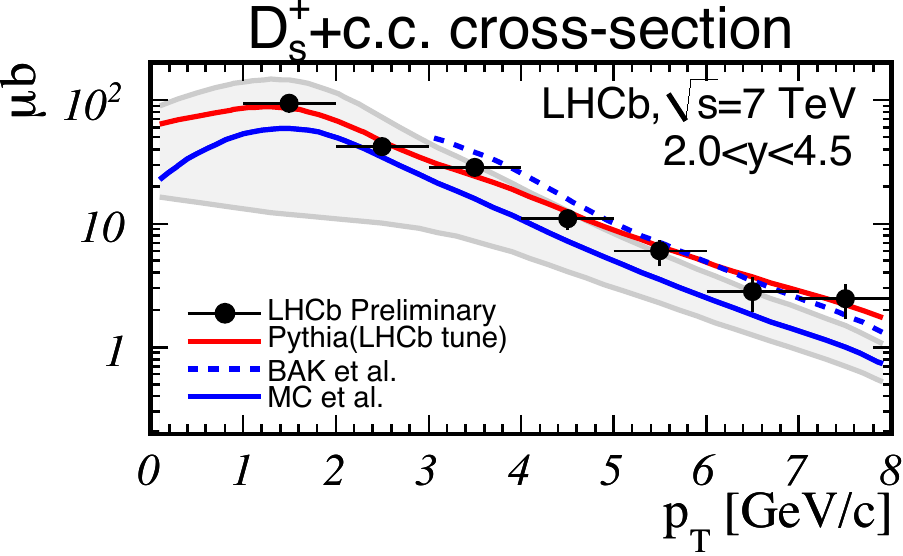}\\
\includegraphics[width=0.67\textwidth]{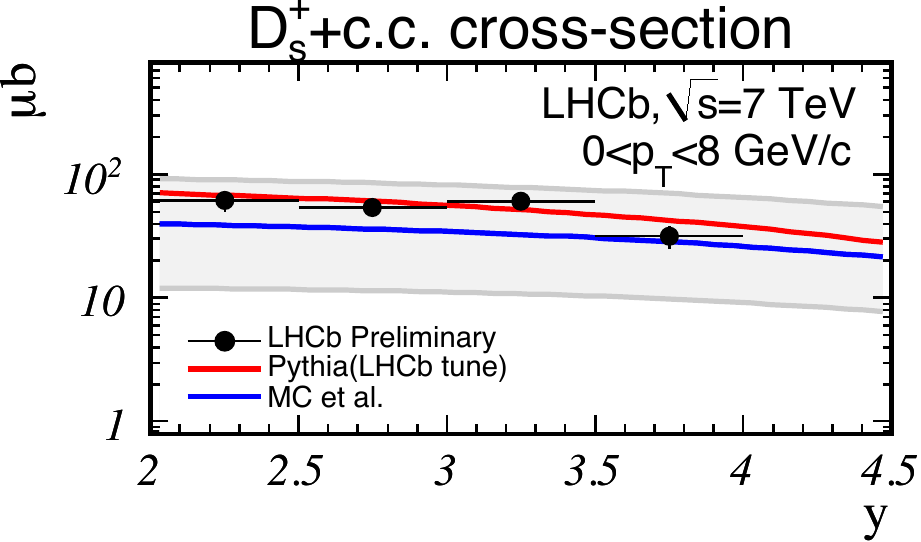}
\caption{Measured \PDsplus cross-sections compared to theoretical prediction. Data are shown as a function of transverse momentum integrated over $2 < y < 4.5$ and as a function of rapidity integrated over $0 < p_T < 8 \gevc$. The error bars show statistical and uncorrelated systematics added in quadrature. In addition there is a global correlated error of $16\%$.}
\label{fig:xsec_ds}
\end{minipage}
\end{figure}

Preliminary results are given in Figures~\ref{fig:xsec_dzero} to~\ref{fig:xsec_ds}.
The data points indicate the integrated cross-section per bin where the vertical error bars include statistical uncertainties and uncorrelated systematic uncertainties.
A correlated systematic uncertainty has to added to the scale of all cross-sections which is detailed in the respective figure captions.

All measured cross-sections are in good agreement with the two theory predictions included in the plots for comparison~\cite{bib:xsec_theo}.
Good agreement is also found with the Pythia~\cite{bib:pythia} prediction using the tuning of the LHCb MC simulation.
This assures good reliability of sensitivity estimates for the physics analyses described below made based on the MC simulation.
More details on these measurements can be found in~\cite{bib:xsec_conf_note}. 

\section{Charm Mixing Measurements}
\label{sec:mixing}
LHCb is currently focussing on the analysis of prompt charm decays.
This ensures very high signal yields, however, adds the complication of the presence of charm decays from long lived, heavier particles (secondary charm).
The separation of prompt and secondary decays is one of the main challenges for all analyses using prompt charm decays.
For time integrated or binned time dependent methods the extraction can be done similarly to the cross-section measurement, i.e. with a fit to the \PD impact parameter distribution.
For unbinned, time-dependent fits, the full time dependence of the \PD impact parameter for prompt and secondary decays has to be modelled and fitted.

A second aspect in common for all time dependent analyses is the necessity for correcting for the lifetime bias caused by cuts in the trigger and offline.
Due to phase space effects these biases do not fully cancel even in measurements of lifetime ratios of different decay modes.
Several data driven methods (see e.g.~\cite{bib:swimming}) are in place to correct for lifetime biases.

The first analysis of charm mixing at LHCb will be performed with the measurement of \ycp.
This observable is measured through the ratio of the flavour averaged lifetime of the $\PDzero\to\PKplus\PKminus$ decay to the lifetime extracted from the decay $\PDzero\to\PKminus\Ppiplus$:
\begin{equation}
\ycp = \frac{\tau(\PDzero\to\PKminus\Ppiplus)}{\tau(\PDzero\to\PKplus\PKminus)} - 1.
\end{equation}
For measuring \ycp both flavour tagged and untagged \PDzero decays are used.
Flavour tagging information is obtained from the charge of the pion in the decay chain $\PDstar^+\to\PDzero\Ppiplus$.
In addition to the information on mixing, CP violation is established by the observation of a difference of \ycp\ and a direct measurement of $y$.

Another promising analysis uses the doubly Cabibbo suppressed decay $\PDzero\to\PKplus\Ppiminus$ (and charge conjugate).
The measurement of the time evolution of this decay gives access to the mixing parameters $x^{\prime2}$ and $y^\prime$, where $x^{\prime}$ and $y^\prime$ are rotated from $x$ and $y$ by the relative strong phase between the doubly Cabibbo suppressed and the Cabibbo favoured decay amplitude.
The low branching ratio of this mode and the tight selection necessary to extract a clean signal require significantly more luminosity than that collected in 2010 in order to make a measurement which improves the current world average.
A sufficient data sample for this measurement is expected to be acquired in 2011.

The same analysis can be performed using the decay $\PDzero\to\PKplus\Ppiminus\Ppizero$.
Due to a low efficiency for reconstructing \Ppizero decays it will require more luminosity to achieve the same sensitivity compared to the $\PDzero\to\PKplus\Ppiminus$ analysis.

A very promising, but also extremely challenging measurement is the time-dependent Dalitz plot analysis of the decay $\PDzero\to\PKs h^+h^-$ ($h=\PK,\Ppi$).
This study gives access as well to the mixing parameters $x$ and $y$ as to the CP violation parameters $\phi$ and $|q/p|$.
To surpass existing measurements in precision this measurement will require the full 2011 data set.

\section{CP Violation in Charm Decays}
\label{sec:cpvmixing}

A direct way of observing CP violation is via the flavour tagged lifetime asymmetry between $\PDzero\to\PKplus\PKminus$ and $\APDzero\to\PKplus\PKminus$ decays:
\begin{equation}
\agamma=\frac{\tau(\PDzero\to\PKplus\PKminus)-\tau(\APDzero\to\PKplus\PKminus)}{\tau(\PDzero\to\PKplus\PKminus)+\tau(\APDzero\to\PKplus\PKminus)}.
\end{equation}
Any non-zero measurement of \agamma represents an unambiguous signal for CP violation in charm mixing.

Both the \ycp\ and the \agamma\ measurements are planned using two independent methods.
One is based on measuring yield ratios in bins of proper time.
This exploits that the lifetime biases largely cancel in the individual bins.

The second method exploits direct, unbinned lifetime measurements.
A method exists to control the absolute lifetime bias using data alone~\cite{bib:swimming}.
This has the advantage of potentially better statistical sensitivity at the cost of more challenging systematics.
A data set to allow a significant improvement of the current world average should be available in the course of 2011.

Another measurement planned is that of T-odd correlations in the decay $\PDzero\to\PKplus\PKminus\Ppiplus\Ppiminus$.
This will be complemented by the search for the rare decay $\PDzero\to\PKplus\PKminus\Pmu^+\Pmu^-$.

\label{sec:cpvdecay}

In the sector of charged charm decays CP violation measurements will be performed using three body final states.
CP violation will be measured using an analysis of the charge asymmetry in bins of Dalitz space~\cite{bib:miranda}.

The Cabibbo favoured decay mode $\PDplus\to\PKminus\Ppiplus\Ppiplus$ will serve as a control mode.
A signal of about $75000$ signal candidates has been measured using the first $128~\nbinv$ of data.
On the same data set clean signals of over $1000$ candidates for the final states $\PKplus\PKminus\Ppiplus$ and $\Ppiplus\Ppiplus\Ppiminus$ each originating from both \PDplus and \PDsplus have been observed.
With the data scheduled to be taken in 2011 it is foreseen to reduce the current uncertainties on CP violation measurements in these channels.

The search for CP violation in the channels $\PDplus\to\PKs\PKplus$ and $\PDplus\to\PKs\Ppiplus$ will complement the measurements of three body final states.
The relatively long lifetime of the $\PKs$ makes a significant fraction of the particles leave the VELO before decaying which complicates the selection of these decays, particularly in the trigger.
The measurements will be performed with data taken in 2011.

\section{Conclusion}
\label{sec:conclusion}
The measurement of the open charm cross-sections of the \PDzero, $\PDstar^+$, \PDplus, and \PDsplus mesons has been presented.
All cross-sections show broad agreement with theory as well as with the tuning of Pythia used to produce LHCb Monte Carlo simulation.
This first analysis of open charm decays underlines the capability of LHCb to make precision charm measurements in the future.
The prospects for performing studies of mixing and CP violation in charm decays are discussed.
Most analyses are expected to yield results improving the current world averages based on the data foreseen to be taken in 2011.

\end{document}

%% file: econfmacros.tex
\def\beq{\begin{equation}}
\def\eeq#1{\label{#1}\end{equation}}
\def\eeqn{\end{equation}}

\def\beqa{\begin{eqnarray}}
\def\eeqa#1{\label{#1}\end{eqnarray}}
\def\eeqan{\end{eqnarray}}

\let\bar=\overbar

\def\Dslash{\not{\hbox{\kern-4pt $D$}}}
\def\dslash{\not{\hbox{\kern-2pt $\del$}}}

\def\msb{{\bar{\ssstyle M \kern -1pt S}}}

